# A NOVEL HIERARCHICAL ANT BASED QOS AWARE INTELLIGENT ROUTING SCHEME FOR MANETS

Debajit Sensarma [1] and Koushik Majumder [1]

[1] Department of Computer Science & Engineering, West Bengal University of Technology, Kolkata, INDIA

***ABSTRACT***

*MANET is a collection of mobile devices with no centralized control and no pre-existing infrastructures. Due to the nodal mobility, supporting QoS during routing in this type of networks is a very challenging task. To tackle this type of overhead many routing algorithms with clustering approach have been proposed. Clustering is an effective method for resource management regarding network performance, routing protocol design, QoS etc. Most of the flat network architecture contains homogeneous capacity of nodes but in real time nodes are with heterogeneous capacity and transmission power. Hierarchical routing provides routing through this kind of heterogeneous nodes. Here, routes can be recorded hierarchically, across clusters to increase routing flexibility. Besides this, it increases scalability and robustness of routes. In this paper, a novel ant based QoS aware routing is proposed on a three level hierarchical cluster based topology in MANET which will be more scalable and efficient compared to flat architecture and will give better throughput.*

## 1. INTRODUCTION

A Mobile Ad Hoc network (MANET) [1] is a dynamically formed wireless network by the mobile nodes. In this network nodes can move randomly. Due to the node mobility network is not stable and topology is not fixed. So, stable routing is a very essential part in MANET. There are three kinds of routing: Proactive, reactive and hybrid. In proactive routing, topology information is distributed proactively; even there is no data to send. So, control overhead increased. To overcome this problem, reactive approach is introduced, where route search is initiated only when needed. But in this routing, there is an initial route discovery delay which is undesirable in many scenarios. So, to overcome these problems, hybrid routing is proposed which is the combination of both the proactive and reactive routing. Alternatively, it is possible to form a cluster of nodes which produces communication hierarchy. There are some advantages of this type of hierarchical routing. Many of the contemporary ad hoc networks are heterogeneous in nature. Mobile devices of this kind of network are equipped with different communication capabilities with respect to frequency band, battery power, radio range, data rate etc. e.g. in military networks, soldiers, tanks and command posts works in different interfaces. So, scalability of this kind of heterogeneous wireless networks is a most important factor. Hierarchical routing makes the protocol more scalable. Flat routing protocols cannot differentiate the mobile nodes with different capacities. Thus, performance of network degrades as the number of mobile nodes with different capacity increases. Furthermore, control overhead, routing overhead much decreases with the hierarchical routing scheme. Hierarchical routing keeps the routing table size smaller in comparison with flat routing scheme. Failure is isolated in hierarchical network topology. So, overhead of route maintenance also decreases.

In this paper, clustering technique is used to organize mobile nodes in small cluster to provide a hierarchical network structure of heterogeneous mobile nodes in MANET. This algorithm takes advantages of Ant Colony Optimization (ACO) [3]. It uses ACO technique for selecting cluster heads. Also intra-cluster routing is done by ACO. It also provides QoS provisioning in each level of hierarchy for efficient and scalable QoS aware route discovery and maintenance. This proposed routing scheme decreases the overhead and improves the overall performance of MANET.
The paper is organized as follows: Section 2 describes the Ant colony optimization. Section 3 describes the related works. In section 4 assumptions of this algorithm is given. Section 5 describes the design of the protocol. Section 6 explains the proposed routing protocol. Section 7 explains the performance analysis. Finally section 8 concludes the paper.

## 2. Ant Colony Optimization

The Ant colony optimization is based on the foraging behavior of ants [3, 11]. When ant started food searching they wonder randomly. When they find the food they return to their colony laying a chemical substance called pheromone. The ants travel randomly exploring all paths but the ants which travel the shortest path reinforce the path with more pheromone. Most of the ant follows the path which has the greater pheromone intensity. This autocatalytic behavior quickly identifies the shortest path.

Some properties characterizes ACO instances for routing problems, they are:

a. In a network where the topology changes dynamically, highly adaptive routing is necessary. Also, in the network without any centralized control, due to node mobility the link can be broken any time and the communication may be lost. If multiple paths exist between source and the destination, one path lost cannot effect the communication, because anyone of the existing paths can be used for routing. ACO provides both the traffic-adaptive and the multipath routing.
b. It is necessary to choose a path for routing which satisfies both the required constraints for routing, for this some previous information are needed and based on the newer and the previous information the path is chosen. In ACO, both the passive and active information are gathered and monitored.
c. ACO uses the stochastic components for routing.
d. ACO does not allow local search estimates to have global impact for the required solution. In ACO no routing information has to transmit to neighbor or all the nodes.
e. ACO does not set paths like other greedy shortest path schemes, at the time of path set up it also taken care of load balancing. So, it taken care of the link quality also.
f. Another important aspect is parameter setting. It is done by ACO in less sensitive way.

Figure 1 illustrates the behavior of ants. A set of ants moves along a straight line from their nest A to a food source B (Figure 1a). At a given moment, an obstacle is put across this way so that side (E) is longer than side (F) (Figure 1b). Now, the ants have to decide which direction they will take: either E or F. The first ones will choose a random direction and will deposit pheromone along their way. The ants taking the way AEB (or AFB), will arrive at the end of the obstacle (depositing more pheromone on their way) before those that take the way AEB (or AFB). So, pheromone intensity of route AFB becomes greater than that of route AEB. So, the ants choose the path AFB (Figure 1c). The ants will then find the shortest way between their nest and the food source.

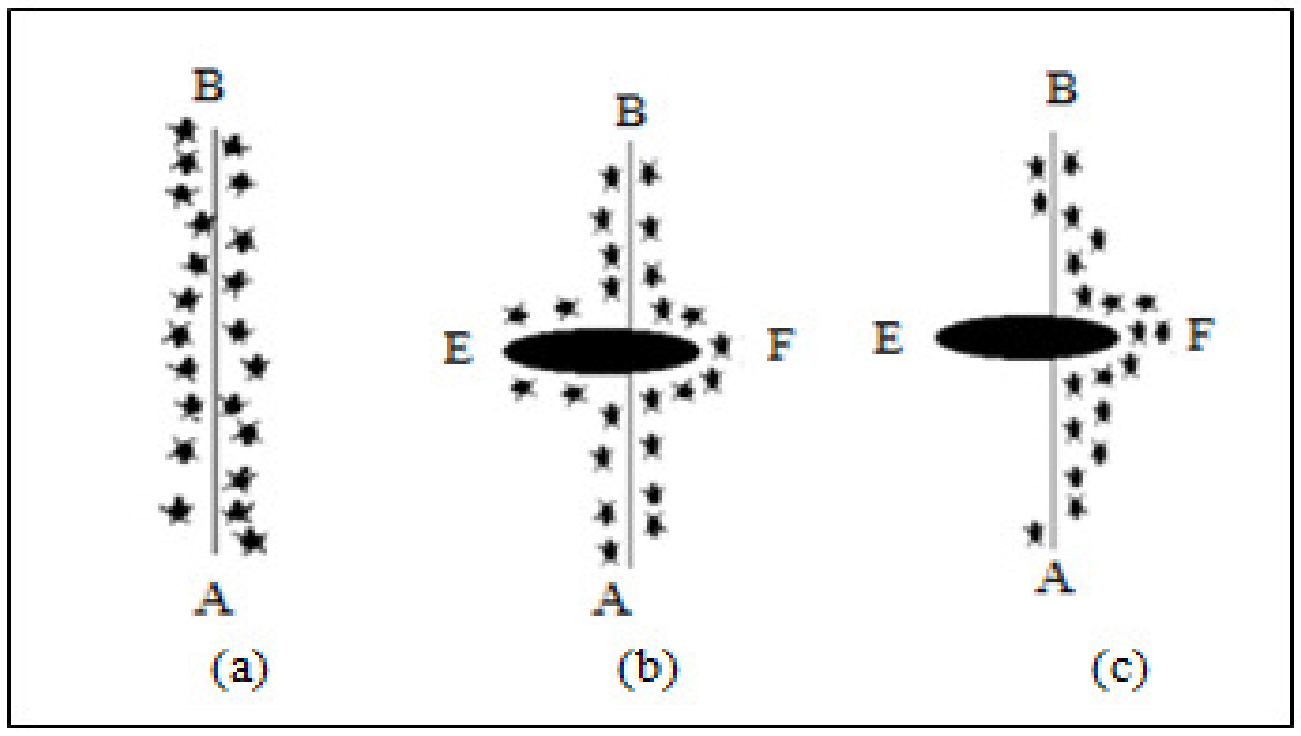

**Fig. 1.** Behaviour of ants for searching the food from A to B

The mechanism of ant colony optimization is described below:

Suppose, an artificial ant deposit a quantity of pheromone represented by $\Delta\tau_{i,j}$ only after completing their route and not in an incremental way during their advancement. This quantity of pheromone is a function of the found route quality.

Pheromone is a volatile substance. An ant changes the amount of pheromone on the path (i, j) when moving from node i to node j as follows:

$$\tau_{i,j} = \sigma.\tau_{i,j} + \Delta\tau_{i,j} \qquad (1)$$

Where $0 < \rho < 1$ and $\rho$ is the pheromone evaporation factor which avoids infinite increment of pheromone which may leads to stagnation of the route.

At one point i, an ant chooses the point j (i.e. to follow the path (i, j)) according to the following probability:

$$P_{i,j} = \frac{(\tau_{ij})^{\alpha}.(\eta_{ij})^{\beta}}{\sum_{i,k\in C}(\tau_{ik})^{\alpha}.(\eta_{ik})^{\beta}} \qquad (2)$$

Where,

$\tau_{i,j}$: is the pheromone intensity on path ( i, j).

$\eta_{i,j}$: is the ant's visibility field on path ( i, j)(an ant assumes that there is food at the end of this path).

$\alpha$ and $\beta$ : are the parameters which control the relative importance of the pheromone intensity compared to ant's visibility field.

C: represents the set of possible paths starting from point i (( i, k ) is a path of C).

Like real pheromone the artificial pheromone decreases over time for fast convergence of pheromone on the edges. This happen in ACO according to the following formula:

$$\tau_{i,j} = (1-q).\ \tau_{i,j} \qquad q\in (0,1] \qquad (3)$$

## 3. RELATED WORKS

Several cluster based hierarchical routing protocols has been proposed. In [4] a two layer heterogeneous network has been proposed. Here, the first layer is formed by the mobile nodes with 802.11 standards and second layer consists of mobile nodes with long range links. It provides the internet access to the lower layer nodes but it does not provide communication between lower layer nodes. The upper layer nodes are gateways to provide the internet access to the lower layer nodes. In [5], an end-to-end QoS aware routing in physically hierarchical Ad hoc network has been proposed. It considers the QoS metrics bandwidth and node speed but other QoS metrics does not taken care of. In [6], HAODV is proposed which is an extension of AODV. Here nodes are heterogeneous in nature and operates on Wi-Fi or Bluetooth. It is a reactive routing, i.e. route discovery initiated only when needed. The advantage is reactive approach helps to reduce the control overhead. But the disadvantage is as there is an initial route discovery delay, so there can be lacks of scalability. As every time number of nodes increases, new route discovery initiated. In [7], a dynamic adaptive routing protocol (DARP) has been proposed. It is same as [6] and suffers from lacks of scalability. In [8], a weight based adaptive clustering has been proposed. It tries to improve routing in heterogeneous MANET using Global Positioning System (GPS). So, by knowing the user mobility pattern accurately, the routing efficiency is increased. But when no GPS is available it is not very useful. In [9], a heterogeneous routing protocol based stable routing has been proposed. It is hybrid clusters based routing protocol and combines AODV and DSDV for intra cluster and inter cluster routing. But disadvantage is, there can be a delay in AODV route discovery process.

There is some advantages and disadvantages of the above hierarchical cluster based routing algorithms. Our proposed routing algorithm takes the advantages of both ant colony optimization and cluster based routing. Here three layer hierarchies are used. Nodes in the lower layer can communicate with each other. Here the QoS constraints delay, bandwidth, energy, link expiration time, hop count are considered for route discovery and battery power, node connectivity, node mobility, distance is used to select the cluster head for increasing network stability. So, here route discovery delay reduces. It is a power aware routing and also scalability increased. Here GPS is not used, so it is useful when GPS is not available.

## 4. ASSUMPTIONS

The following assumptions are taken into consideration to construct the protocol:

1. Here, a three level hierarchical cluster model is considered.
2. Ant Colony Optimization is used here for cluster head selection. Cluster heads are selected based upon the weight value of the nodes. In each layer same procedure is followed. Each cluster head can only have one hop neighbours and neighbours can communicate with each other.
3. A node can communicate maximum with three levels based on the transmission power and battery power. Level-0 nodes can communicate with only level-0 nodes. Level-1 nodes can communicate with level-1 and level-0 nodes and level-2 nodes can communicate with all three level nodes.
4. Here, hierarchical addressing is used for each node and based on the transmission power nodes are represented with different shapes. Details of this presentation are depicted in section 5.1.
5. In each level QoS constrains are taken care of and Ant Colony Optimization is used for both intra and inter cluster routing.

# 5. DESIGN OF PROPOSED PROTOCOL

## 5.1 LOGICAL TOPOLOGY LEVELS

Fig. 2 illustrates the network architecture of the proposed routing protocol. The nodes are organized in multiple topology levels based on the capacity. The nodes at topology Level-0 are represented by white circle and they are equipped with only one interface with limited data rate. Lavel-1nodes are represented by dark circle and they are equipped with two interfaces, one is the interface with level-0 nodes, i.e. they can communicate with level-0 nodes via wireless channel and next is they can relay messages to other level-1 nodes using channel different from level-0 topology having longer transmission range. Topology level-2 nodes are represented by triangles and have three interfaces. They are capable of communicating with level-0 and level-1 nodes and other level-2 nodes via high speed point to point direct wireless links.

The architecture is designed as follows- Each clusters are leveled with ‘C’ followed by a topology level at which the cluster is formed, followed by cluster head id of that cluster. For example, C0.A1 denotes level-0 cluster having A1 as cluster head. The nodes having single interface are denoted by white circles and represented by single digit (1, 2, 3 etc.). These nodes are formed only in bottom level. Multiple interface nodes are represented by node’s name (e.g. A1, A2 etc.) followed by a digit denoting the node’s interface where it operates. Nodes with interfaces indicated by triangles can operate in three levels (e.g. A1.2, A1.1, and A1.0) and nodes denoted by dark circles can communicate with lower levels (e.g. A3.1, A3.0).

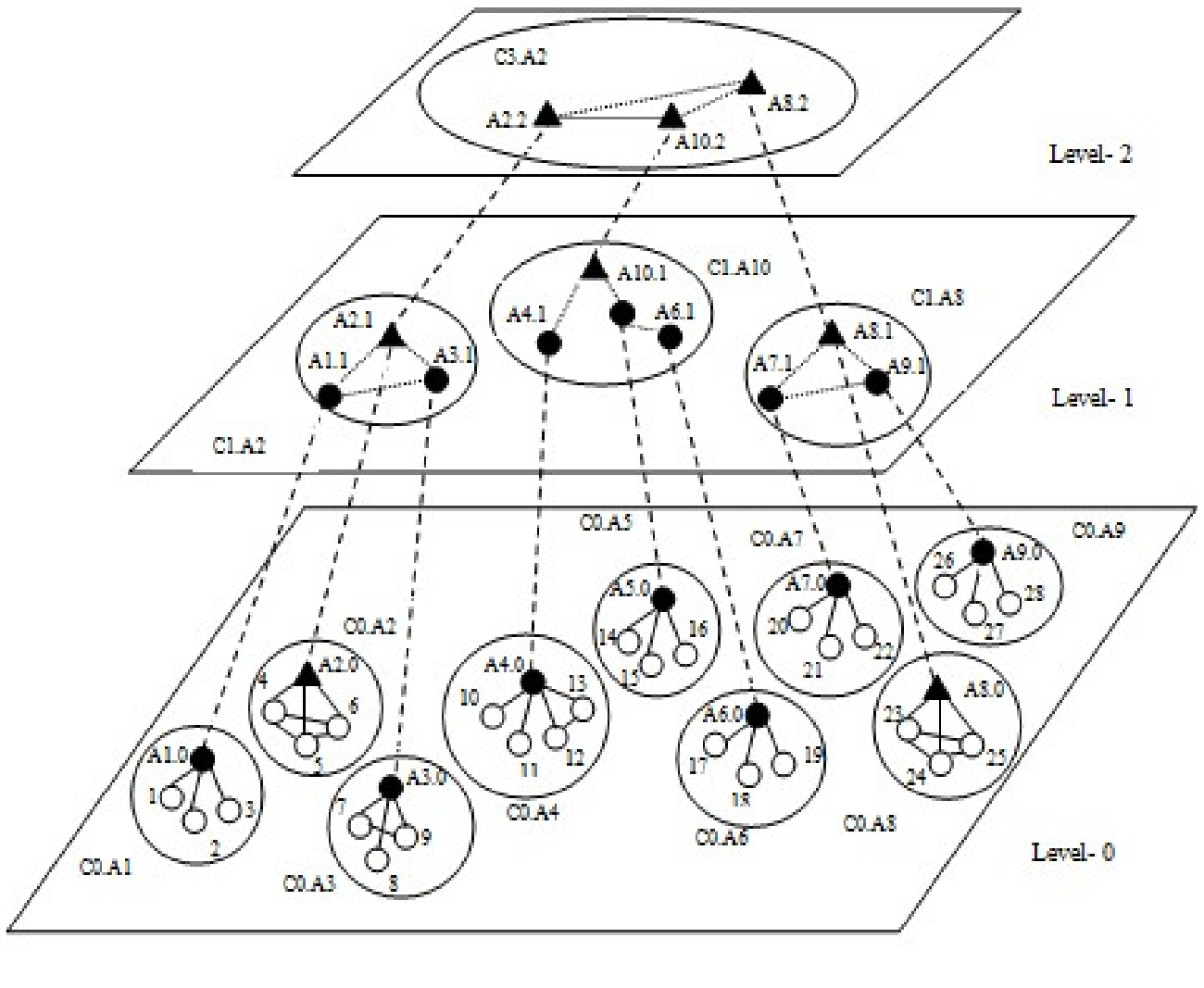

○ Member node (node with 1 interface)

● Cluster head (node with 2 interfaces)

▲ Cluster head (node with 3 interfaces)

- - - Connection between inter label clusters

······· Connection between intra label clusters

**Fig.2.** Architecture of proposed three layer hierarchical structure

### 5.2 Cluster Head Selection and Cluster Formation

In cluster head selection process, at first, nodes find their neighbour according to range of each other. Here Ant Colony Optimization is used to select a cluster head. At first, cluster of nodes are formed. Then for each cluster, Cluster head is selected based on the weight function, pheromone value, probability. According to [10] cluster head can be selected based on combined weight metric of the node. But, there is a threshold on weight value for selecting a node as a cluster head. In this procedure, cluster head is selected with following weight factors:

- **Battery power:** The remaining power of the node.
- **Node Connectivity:** The Number of nodes that can directly communicate with the given node within the transmission range of each other.
- **Node Mobility:** Average running speed of the node. Lesser the mobility, higher the probability of selection.
- **Distance:** Sum of distance of the node from all its neighbours.

In this cluster head selection procedure each cluster head can have only one hop neighbour and each neighbour can be connected with each other.

The algorithm is described below:

---

**Algorithm 1: Weight Calculation Algorithm.**

---

**Begin**

1. Find connectivity 'c' for each node which is the number of neighbours of each node. Find the remaining energy, 'e' for each node.
2. Compute the mobility m for each node which is the running.
3. Compute the sum of distances d with all its neighbours for each node.
4. Calculate the combined weights $weight_i$ as

    $weight_i = w1*c_i + w2*e_i - w3*m_i + w4*d_i$

    Here w1+w2+w3+w4=1.

    $c_i$ = Connectivity of node i. i.e. number of nods that can directly communicate with the given node within the transmission range of each other.

    $e_i$ = Battery power of node i. i.e. remaining power of that node.

    $m_i$ = Mobility of node i.

    $d_i$ = Distance of each node from i. i.e. sum of distance of the node from all its neighbours.

**End**

---

**Algorithm 2: Cluster Head Selection Algorithm.**

---

**Begin**

1. Each node finds its neighbours and builds its neighbourhood table.
2. A set of clusters are formed from nodes and its neighbours.
3. Each node calculates its weight by calling the weight calculation algorithm given above.
4. At first, in a cluster, a node is selected as a cluster head randomly.
5. In each iteration, a node is selected as a cluster head based on probability of the neighbour nodes. The probability of each node to be selected as cluster head is:

$$P_i = \frac{\tau_i}{\sum_{k=1}^{n} \tau_k} \quad (1)$$

6. Each time a node is selected as the cluster head, the pheromone value is updated according to the following formula:

$$\tau_i = (1-\rho)\tau_i + \rho(\text{weight}_i) \quad (2)$$

Here $\rho$ is the pheromone evaporation factor ($0<\rho<1$).

7. Continue step 5-6 for each node in the cluster until all the nodes in the network are covered (i.e. a node is a cluster head or falls within the range of an existing cluster head).
8. The node, whose weight value and the pheromone value is greater than its neighbour and greater than threshold, is selected as cluster head and it sends message to all its neighbours.
9. On receiving a message, all neighbour nodes unicast an acknowledgment message to the selected cluster head.
10. For each cluster continue the steps 3-9.

**End**

In level-0 after all nodes joins the cluster, level-0 cluster heads broadcast the cluster information across the network. When one cluster head receives the information from other cluster head, it sends acknowledgement to the source cluster head and a link is established. Level-1 and level-2 cluster forms in same procedure.

In this routing after formation of hierarchical cluster two cases can occur:

Case 1: During the data transmission, cluster head weight value decreases time by time because of power dissipation. If the weight value of a cluster head becomes lower than the specified threshold, cluster head sends a control message to all its neighbour informing them to start a new cluster head selection procedure.
Case 2: If a new node arrives with greater power, then also a new cluster head selection procedure started.

## 6. PROPOSED ROUTING ALGORITHM

This is a three level hierarchical cluster based routing scheme which utilizes the Ant Colony Optimization. This is also a QoS aware routing with parameters: delay, bandwidth, energy and link expiration time. It takes care of QoS constrains in each layer. Here 5 kinds of packets are used. Route_Ant is used by a node when it wants to know if the destination is a member of its cluster head or not. Request Knave_Ant and Reply Knave_Ant are used for intra cluster routing. Request King_Ant and Reply King_Ant are used for inter cluster routing. This algorithm has two phases: Route discovery phase and route maintenance phase.

### 6.1. MATHEMATICAL MODEL

For mathematical analysis MANET is represented by a connected undirected graph. Let G (V, E) represents the mobile ad hoc network. Here V denotes the set of network nodes and E denotes the set of bidirectional links. QoS metrics with respect to each link $e \in E$ is delay (e), bandwidth (e), link expiration time (e). With respected to node $n \in V$, it is delay (n), energy (n). Another QoS metric considered here is hop count. It is important because multiple hops are used for data transmission in MANET. So, it is necessary to find paths with minimum hops. The main motivation of this proposed algorithm is to find path from source to destination which will satisfy the QoS requirements such as delay, bandwidth, energy, link expiration time.

Let, path (i, j) or R is entire path from node i to j where QoS constraints have to satisfied.

From an arbitrary node i to an arbitrary node j, delay, bandwidth, energy, link expiration time and hop count is calculated as-

delay (path (i, j )) or D (R) = $\sum_{e \in P(i,j)} delay(e) + \sum_{n \in P(i,j)} delay(n)$

where, delay (e) is the transmission and propagation delay of the path(i,j) and delay (n) is the processing and queuing delay of node 'n' on path(i, j).

bandwidth(path(i,j)) or B(R)= $\min_{e \in P(i,j)}$ {bandwidth(e)}

where, bandwidth (e) is the available bandwidth of that link on path(i, j).

link expiration time or T(R) = $\min_{e \in P(i,j)}$ { link expiration time (e)}

where, link expiration time (e) is the expiration time of a link in route R.

energy (path (i, j)) or E (R) = $\min_{n \in P(i,j)}$ { energy (n)}

where, energy (n) is the residual energy of node 'n' on path(i, j).

hop count (path (i, j)) or HC (R) = Number of nodes in the path.

### 6.2. Calculation of Pheromone

Ant deposits pheromone during traversal of the link for finding a route. The quantity of pheromone it deposited on each link (i, j) along the route R is noted by $\Delta\tau_{i,j}$ and it is a function of global quality of route R. It is expressed by the following equation-

$$\Delta\tau_{i,j} = \frac{B(R)^{\lambda_B} + E(R)^{\lambda_E} + T(R)^{\lambda_T}}{D(R)^{\lambda_D} + HC(R)^{\lambda_{HC}}} \quad (3)$$

Here $\lambda_B$, $\lambda_E$, $\lambda_D$, $\lambda_{HC}$ and $\lambda_T$ are the weight factors which indicate the relative significance of the QoS parameters during pheromone update on path (i, j). The quantity of the deposited pheromone is defined only after finding the route.

### 6.3. Calculation of Path Preference Probability

Path Preference Probability is calculated in each intermediate node as well as source node upon receiving of Reply Knave_Ant or Reply King_Ant.

Suppose current node i receives Reply Knave_Ant or Reply King_Ant from node j for destination d, then the Path Preference Probability is calculated as-

$$P_{ijd} = \frac{[\tau_{ij}]^{\alpha 1}.[D_{ijd}]^{\alpha 2}.[\eta_{ijd}]^{\alpha 3}.[B_{ijd}]^{\alpha 4}.[E_{ijd}]^{\alpha 5}[T_{ijd}]^{\alpha 6}}{\sum_{k \in N_i}[\tau_{ik}]^{\alpha 1}.[D_{ikd}]^{\alpha 2}.[\eta_{ikd}]^{\alpha 3}.[B_{ikd}]^{\alpha 4}.[E_{ikd}]^{\alpha 5}[T_{ijd}]^{\alpha 6}} \quad (4)$$

Here α1, α2, α3, α4, α5 and α6 are the tunable parameters which control the relative weights of pheromone trails, hop count, bandwidth, energy and link expiration time respectively.

$N_i$ is the set of neighbors of i and k is the neighbor node of i through which a path to destination is known.

The relative metrics are calculated from source i to destination d via j as-

$$D_{ijd} = \frac{1}{delay(path(i,d))}$$

$$\eta_{ijd} = \frac{1}{hopcount(path(i,d))}$$

$B_{ijd}$= bandwidth (path(i,d))
$E_{ijd}$ = energy (path (i,d))
$T_{ijd}$ = link expiration time (path (i,d))

Now, source as well as neighbors has multiple paths from source to destination. The path with higher Path Preference Probability is selected for the data transmission.

## 6.4. Packet Formats in the Proposed Routing

**Route_Ant:**

This type of packet is used to identify the information about whether a route to a particular node exists or not. Here 'Flag' field is used to notify the existence of a valid route. In fig.3.the packet format of Route_Ant is shown.

0 11 21 31

| Source Node id | Destination Node id | Flag |
|---|---|---|

**Fig.3.** Route_Ant packet format

**Request Knave_Ant:**

This packet is used in route discovery phase of intra cluster routing. It contains request starting time, available bandwidth, member source id, member destination id, stack of visited node addresses.

0 2 17 31

| Pkt_Type = Request Knave_Ant | Req_Starting Time | Bandwidth |
|---|---|---|
| Member Source ID | | |
| Member Destination ID | | |
| Stack of nodes visited<br>: | | |

**Fig.4.** Request Knave_Ant Packet format

### Reply Knave_Ant:

This packet also is used in route discovery phase of intra cluster routing. It contains hop count, delay, energy, link expiration time, available bandwidth, member destination id, member source id, stack of node addresses to be visited.

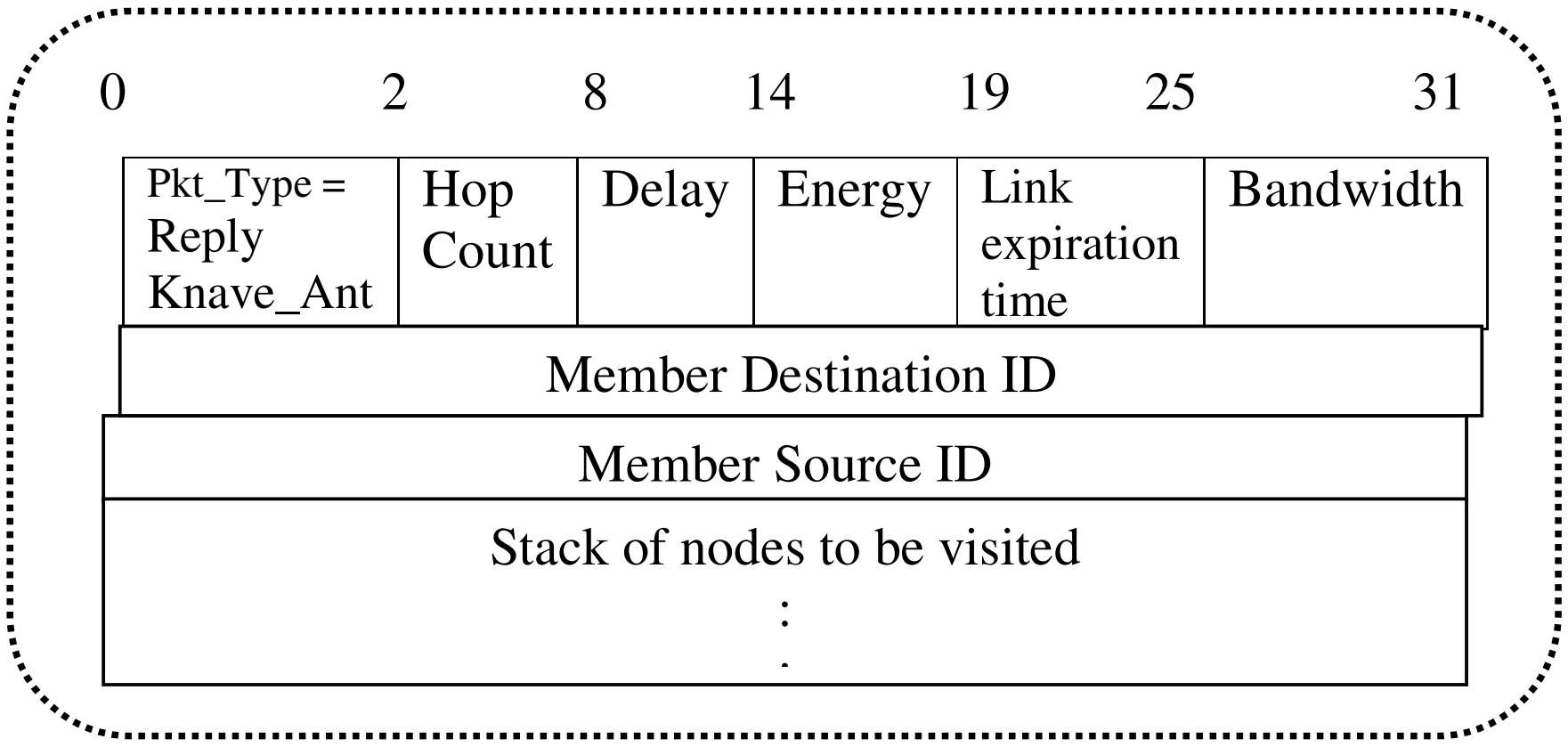

**Fig.5.** Reply Knave_Ant Packet format

### Request King_Ant:

This packet is used in route discovery phase of inter cluster routing. It contains request starting time, available bandwidth, Cluster head source id, Cluster head destination id, stack of visited node addresses.

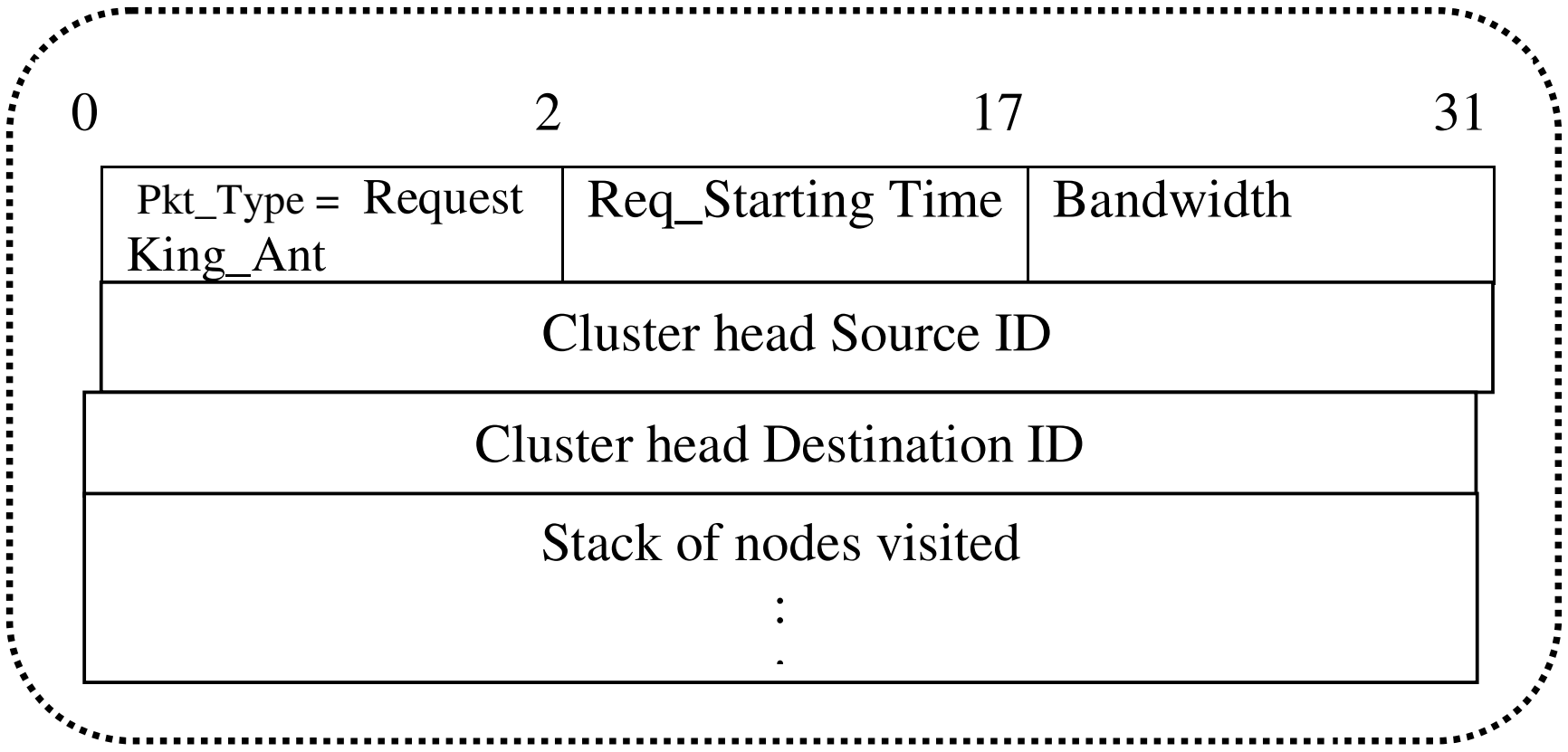

**Fig.6.** Request King_Ant Packet format

### Reply King_Ant:

This packet also is used in route discovery phase of inter cluster routing. It contains hop count, delay, energy, link expiration time, available bandwidth, Cluster head destination id, Cluster head source id, stack of node addresses to be visited.

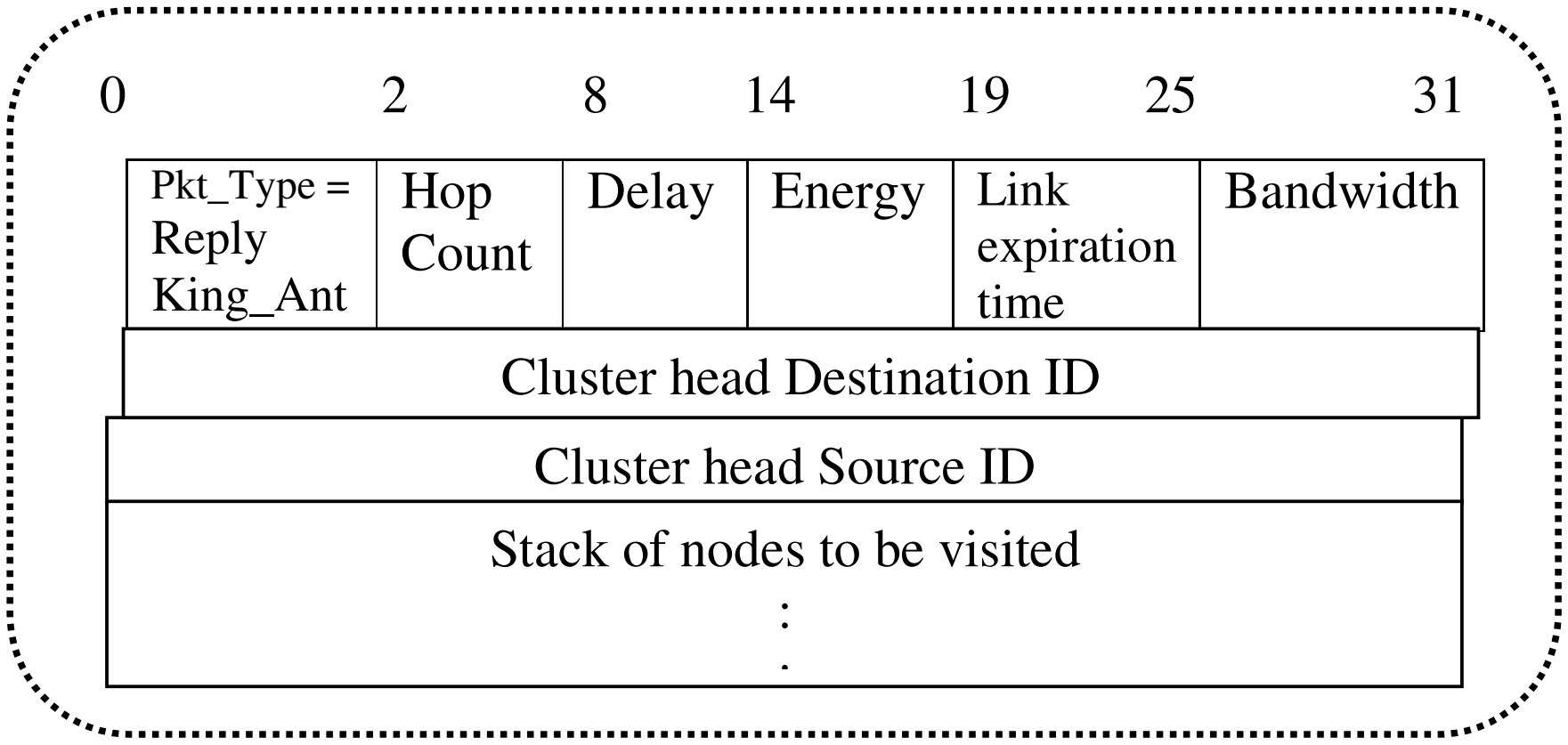

**Fig.7.** Reply King_Ant Packet format

### 6.5. ROUTING IN PROPOSED HIERARCHICAL NETWORK

---

**Algorithm 3: Route Discovery Phase**

---

**BEGIN**

Suppose Source S wants to communicate with destination D.

/* Intra-cluster routing */

**Step 1:** S searches in its neighbour table to see if D is a neighbour of S.

**Step 2:** If S finds that D is its own neighbour, then it directly sends data packets to D.

**Step 3:** If S finds that D is not its neighbour, it unicast the Route_Ant packet to its cluster head (i.e. CH(S)).

**Step 4:** CH(S) searches in its member table to see if D is a member of it.

**Step5:** If CH(S) finds that D is its own member, then it unicast the Route_Ant to S by setting flag=1.

**Step 6:** After receiving Route_Ant from cluster head, S starts finding route to D which will satisfy the required QoS constrains delay, bandwidth, energy, link expiration time and hop count.

**Step 7:** It first consults with its route cache, if an unexpired route exits to D, and then the route with higher path preference probability is used for transmitting data.

**Step 8:** If cache contains no unexpired route, S initiates a Request Knave_Ant to the destination through all its neighbours.

**Step 9:** While travelling to the destination the Request Knave_Ant collects delay, bandwidth of each link and energy, link expiration time of each node.

**Step 10:** When the Request Knave_Ant reaches the destination it will converted to Reply Knave_Ant and forwarded towards the original source. It will take same path as the Request Knave_Ant but in reverse direction.

**Step 11:** For every Reply Knave_Ant reaching an intermediate node or source node, node can find the delay, bandwidth, energy and link expiration time to the destination and node calculate Path Preference Probability. If it is better than the requirements, and then the path is accepted and stored in the memory.

**Step 12:** The path with higher Path Preference Probability will be considered as best path and will be selected for transmitting data.

/* Inter-cluster routing in the same network or region */

**Step 13:** IF D is not a member of CH(S), it will unicast Request_Ant to its cluster head (CH (CH(S))) in level-1hierarchy.
**Step 14:** If CH (CH(S))) finds D in its own member table but not in its cluster, as well as D is in same network or region, then it will unicast the Route_Ant to the CH(S) and will set flag=1.
**Step 15:** After receiving Route_Ant from cluster head, CH(S) starts finding route to cluster head of D (i.e. CH (D)), which will also satisfies the QoS requirements.
**Step 16:** At first, it consults with its route cache, if unexpired paths exists to CH (D)), then the path with better Path Preference Probability is selected for data transmission.
**Step 17:** If cache contains no unexpired route, then CH(S) initiates a Request King_Ant to destination through all its neighbour.
**Step 18:** The Request King_Ant also collects the delay, bandwidth, energy, link expiration time and hop count of each links and each nodes respectively.
**Step 19:** When Request King_Ant reaches the destination cluster head, it will be converted to Reply King_Ant and also forwarded to the CH(S) by the same path as Request King_Ant but in reverse direction.
**Step 20:** Intermediate cluster nodes calculate the Path Preference Probability and if the Path Preference Probability is better than requirements, then it is stored in the node's cache.
**Step 21:** The path with higher Path Preference Probability is selected for data transmission and the data is transmitted directly to the destination through its cluster head (CH (D)).
**Step 22:** If CH (CH(S)) finds D in its own member tables the member of its cluster, then the communication starts directly through that cluster head [ i.e. CH CH(S) have at least 2 interfaces].

/* Inter-cluster routing in the different network or region*/
**Step 23:** If CH (CH(S)) finds D in its member table but D is in another network, then it will forward the Request Knave_Ant packet to the corresponding cluster member node with required QoS constraints in level-2. Then the member node unicast the request to the desired destination's cluster head and finally when the Request King_Ant packet reaches the destination, it sends the Reply King_Ant packet to the original source by the same path but in reverse order.
**Step 24:** If CH (CH(S)) does not find D in its member table, it unicast Request packet to cluster head (i.e. CH (CH (CH(S)))) and if D is in its member list, it sets the flag to 1 and sends to CH(CH(S)) and route discovery continues like previous process.

**END**

---

---

**Algorithm 4: Route Maintenance Phase**

---

There are three cases in maintenance: single node leaving the cluster, single node joining the cluster, the whole cluster moving together.

**BEGIN**

**Maintenance of level-0 cluster.**

**Case 1:** Leaving of a node in level-0 cluster.
Cluster head periodically will send the beacon to the member node and the member nodes send acknowledgement to the cluster head. If a cluster head does not receive any reply from its member within a specific period of time, then it can detect leaving of member from the cluster. Again, if the cluster member cannot receive any beacon from the cluster head, then it can detect leaving of its cluster head from the cluster.

**Case 1.1:** If an internal node moves out of a cluster, then the cluster head deletes the member node entry from the member table.

**Case 1.2:** If a cluster head moves out of a cluster, nodes that do not belong to any cluster started the cluster head selection procedure.

**Case 2:** Joining of a node in level-0 cluster.

When a cluster head starts receiving an acknowledgement from a new node, then it adds this new member node to the member table.

**Case 3:** The whole cluster moving together.

When a cluster head comes in the transmission range of another cluster head at level-0, then a new cluster head selection procedure started.

**Maintenance of level-1 cluster.**

**Case 4:** Level-1 change caused by one node leaving level-0 cluster.

**Case 4.1:** Leaving node is the level-1 cluster head.

When level-0 cluster head cannot get any beacon from its cluster head, then a new cluster head selection procedure started in level-1 as well as in level-0 cluster.

**Case 4.2:** Leaving node is not the level-1 cluster head.

When level-1 cluster head cannot find any beacon from a member node which is a cluster head of level-0 cluster, it deletes its corresponding node entry from its member table and a new cluster selection procedure starts at level-0.

**Case 5:** Level-1 change caused by new cluster head joining in level-0 cluster.

When a new cluster head selected in level-0, it is added to the member table of the cluster head of level-1 which is in the same network or region.

**Maintenance of level-2 cluster.**

**Case 6:** Level-2 change caused by one node leaving level-1 cluster.

**Case 6.1:** Leaving node is the level-2 cluster head.

In this case a new cluster head selection procedure is invoked. Here, if the leaving node has two interfaces (i.e. level-2 and level-1), then new cluster head selection procedure is occur in level-2 cluster and level-1. But if it has 3 interfaces, then new cluster head selection procedure occurs in three levels.

**Case 6.2:** Leaving node is not the level-2 cluster head.

In this case, the corresponding node entry is removed from the cluster head member table. Here, if the leaving node has two interfaces (i.e. lavel-2 and level-1), then new cluster head selection procedure is occur only in level-1 cluster and if it has 3 interfaces, then new cluster head selection procedure occurs in level-2 and leve1-1 cluster.

**Case 7:** Level-2 change caused by new cluster head joining in level-1 cluster.

When a new cluster head selected in level-1, it is added to the member table of the cluster head of level-2 cluster.

**END**

## 7. PERFORMANCE ANALYSIS

The proposed scheme is a hierarchical Ant based routing algorithm. It is suitable for the heterogeneous network where nods have difference transmission power and different characteristics and with no centralized control. It takes the advantages of both clustering and Ant Colony Optimization technique.

This algorithm is based on three layer hierarchical structure, where each node has different transmission power and they are different in nature. Here cluster heads are selected based upon the node connectivity, node mobility, transmission power or battery power and distance factor to provide a stable network which helps in increase of routing stability among heterogeneous nature of nodes. It is very essential for real time applications. Besides this, scalability also increased and delay decreased, because every time route discovery is not initiated with increasing number of nodes. Only when node with greater power comes, that time a new cluster head selection procedure started. Here Ant colony optimization is used, so no routing table is transmitted to neighbors. Thus it decreases the routing overhead. Here, in route discovery phase both intra cluster and inter cluster routing is considered with the QoS metrics: delay, bandwidth, energy, link expiration time and hop count. So, it is a power aware routing. Here, also maintenance in each layer is considered separately to provide better network throughput.

## 8. CONCLUSION & FUTURE WORKS

As MANET is dynamic in nature, so QoS provisioning is very difficult in this network. In this paper we proposed an Ant based hierarchical on-demand routing. It is a three level cluster based routing algorithm. It takes the advantages of both the ant colony optimization and cluster architecture. It is also a power efficient routing. Node's transmission power plays a very crucial role for increasing routing stability. Unlike other routings, QoS constrains are taken care of each layer. So, it is very efficient for real time communication with heterogeneous nature of nodes. Scalability also increases in this routing scheme. Here, an efficient cluster formation is used for handling the increased number of nodes. Besides this, it emphasize on cluster maintenance for reducing the overhead and delays of the network.

In future we will simulate this protocol and will compare it with other hierarchical routing algorithms. Also we will take new QoS metrics to provide better throughput for real time communication among the heterogeneous nature of nodes.

## ACKNOWLEDGEMENTS

The authors would like to thank West Bengal University Technology, West Bengal, India for the supports and facilities provided to carry out this research. The authors also thank the reviewers for their constructive and helpful comments.

.

## Authors

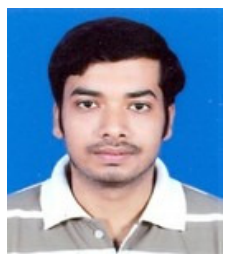

Debajit Sensarma has received his B.Sc. degree in Computer Science in the year 2009 from university of Calcutta, Kolkata, India and M.Sc. degree in computer Science with top rank in the University, in the year 2011 from West Bengal State University, Kolkata, India. He has been awarded the INSPIRE Fellowship by the Department of Science & Technology, New Delhi. He obtained his M.Tech. degree in Computer Science and engineering from West Bengal University of Technology, Kolkata, India, in the year 2013. He is now pursuing his PhD degree from the department of Computer Science and Engineering, University of Calcutta, Kolkata, India. He has published several papers in International journals and conferences.

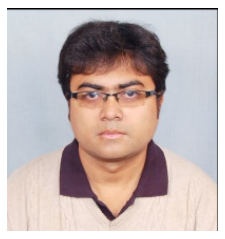

Koushik Majumder has received his B.Tech and M.Tech degrees in Computer Science and Engineering and Information Technology in the year 2003 and 2005 respectively from University of Calcutta, Kolkata, India. He obtained his PhD degree in the field of Mobile Ad Hoc Networking in 2012 from Jadavpur University, Kolkata, India. Before coming to the teaching profession he has worked in reputed international software organizations like Tata Consultancy Services and Cognizant Technology Solutions. He is presently working as an Assistant Professor in the Dept. of Computer Science & Engineering in West Bengal University of Technology, Kolkata, India He has published several papers in International and National level journals and conferences. He is a Senior Member, IEEE.